\documentclass[12pt]{article}
\newcommand{\be}{\begin{equation}}
\newcommand{\ee}{\end{equation}}
\newcommand{\bqa}{\begin{eqnarray}}
\newcommand{\eqa}{\end{eqnarray}}

\begin{document}
\begin{center}
{\LARGE A note on possible interpretations for the $D_{SJ}^+(2632)$ observed by SELEX}\\[0.8cm]
{\large Kuang-Ta Chao$^{(a,b)}$}\\[0.5cm]
{\footnotesize (a)~Department of Physics, Peking University,
 Beijing 100871, People's Republic of China}\\
{\footnotesize (b)~China Center of Advanced Science and Technology
(World Laboratory), Beijing 100080, People's Republic of China}
\end{center}
\vspace{0.5cm}

\begin{abstract}
We suggest some possible interpretations for the $D_{SJ}^+(2632)$
observed by SELEX. The $D_{SJ}^+(2632)$ could be the first radial
excitation of the $1^-$ ground state $\rm{D_s}^*$(2112), and its
unusual decay patten might be hopefully explained by the node
structure of the wave functions. In addition, the $D_{SJ}^+(2632)$
could also be a $c\bar sg$ hybrid state or a $(cs)_{3^*}-(\bar
s\bar s)_{3}$ (diquark-antidiquark) bound state.
\vspace{1cm}\\

\end{abstract}

\vspace{1cm}
Very recently the SELEX Collaboration has reported
the first observation of a charm-strange meson $D_{SJ}^+(2632)$ in
the charm hadro-production experiment E781 at
Fermilab\cite{selex}. The $D_{SJ}^+(2632)$ is observed in the
${D_{s}^{+} \eta}$ decay channel with a significance of
$7.2\sigma$ and in the $D^0K^+$ decay channel with a significance
of $5.3\sigma$. The mass and width of this state are found to be
$M=2632.6\pm 1.6 MeV$ and $\Gamma<17 MeV$ (at $90\%$ confidence
level). This narrow state has a rather unusual decay patten that
it is dominated by the $D_{s}^{+} \eta$ decay mode with a very
small ratio $R=\Gamma(D^0K^+)/\Gamma(D_{s}^{+} \eta)=0.16\pm
0.06$.

For the $c\bar s$ mesons, aside from the ground state doublet
[$\rm{D_s}$(1969), $\rm{D_s}^*$(2112)] of [$0^-,1^-$] states, as
well as the $D_{s1}(2536), D_{sJ}(2573)$, which probably
correspond to the $J^P=1^+,2^+$ states being the P-wave
excitations with the light quark angular momentum $j=3/2$, in 2003
the BaBar collaboration reported the first observation of a
massive, narrow charm-strange meson $D_{s0}(2317)$ below the
$\rm{D K}$ threshold~\cite{babar}. CLEO~\cite{cleo} and
BELLE~\cite{belle} subsequently confirmed the existence of
$D_{sJ}(2317)$, and further found another narrow higher-lying
state $D_{sJ}(2460)$. The finding of [$D_{sJ}(2317)$,
$D_{sJ}(2460)$] has stimulated many theoretical explanations for
these two states\cite{th}. Most likely, they may correspond to the
$J^P=0^+,1^+$ states being the P-wave excitations with the light
quark angular momentum $j=1/2$. In particular, in the chiral
models for the heavy-light mesons, parity doublers are predicted
to have the same mass splittings, and the ground state doublet
[$\rm{D_s}$(1969), $\rm{D_s}^*$(2112)] of [$0^-,1^-$] states would
have a parity-partner [$0^+,1^+$] pair, which are very likely to
be the observed [$D_{sJ}(2317)$,
$D_{sJ}(2460)$]\cite{bardeen,nowak}. In 2004, the new state
$D_{SJ}^+(2632)$ observed by SELEX with unusual decay modes will
certainly be interesting in finding its own place in the
heavy-light systems.

In the following we will present some discussions on the possible
interpretations for the $D_{SJ}^+(2632)$ state.

\vspace{0.5cm}
 (1). The $D_{SJ}^+(2632)$ could be the first radial
excitation of the $1^-$ ground state $\rm{D_s}^*$(2112).
\vspace{0.5cm}

 The $D_{SJ}^+(2632)$ can decay into two psedoscalar
mesons $D_{s}^{+} \eta$ or $D^0K^+$, so it should have
$J^P=0^+,1^-,2^+,...$. For the vector meson, the mass difference
between the ground state and its first radial excitation ranges
between 550-650 MeV. This can be seen from the observed $2S$ and
$1S$ mesons. For instance, the $2S$ and $1S$ states are found to
be, e.g., $\omega(1420)$ and $\omega(782)$ for the light-light
mesons; $\psi(3686)$ and $J/\psi(3097)$ for charmonium; and
$\Upsilon(10023)$ and $\Upsilon(9460)$ for bottomonium. The mass
difference between $D_{SJ}^+(2632)$ and $\rm{D_s}^*$(2112) is 520
MeV, it is marginal but still acceptable for the excitation energy
of the $2S$ heavy-light vector mesons, considering the fact that
the P-wave excitation energy of $D_{sJ}(2317)$ is considerably
smaller than that for the observed light-light and heavy-heavy
mesons and the conventional potential model calculations.

As the radial excitation of the $1^-$ ground state
$\rm{D_s}^*$(2112), the $D_{SJ}^+(2632)$ could have a small decay
width and unusual decay modes due to the node structure in its
wave function. For the OZI allowed hadronic decay, the decay
amplitude is related to the overlap integral of the wave functions
of the initial and final state hadrons, and therefore is sensitive
to the node structure of the wave functions: the sign-changing
wave function of the radially excited states may result in
substantial suppression for the decay rates of certain modes. The
unusual experimental result in charmonium spectrum that the
$\psi(4030)(=\psi(3S))$ has a dominant decay amplitude to the
$D^*\bar D^*$ mode (with a very small $Q$-value) over the $D\bar
D$ mode (with a much larger $Q$-value) might be explained by the
observation\cite{yao} that the nodes of the wave function and the
different $Q$-value in each of these decays allow to understand
the failure of the simple phase space argument. Indeed, the nodes
of the wave functions lead to existence of zeros of the decay
amplitude in the momentum of the decay products, which are
responsible for the suppression of certain decay modes. The widths
of the $\psi(4414)(=\psi(4S)$) decaying into all ground state
charmed meson pairs were estimated to be only about 15~MeV,
despite of the very large phase space\cite{yao}. Although in
\cite{yao} the quark pair creation model (i.e. the $^3P_0$ model)
was used we expect the qualitative features obtained there should
hold regardless which specific model for describing the quark pair
creation was used (see also\cite{eichten} for the treatment in the
Cornell model).

If the above features also hold for the decays of the radially
excited heavy-light mesons, it would be not impossible to
understand why the $D_{SJ}^+(2632)$ could have a narrow width
(say, of order 10~MeV) and could even have the decay mode
${D_{s}^{+} \eta}$ (with a smaller $Q$-value and a $s\bar s$ quark
pair creation) dominating over the decay mode $D^0K^+$ (with a
larger $Q$-value and a $u\bar u$ quark pair creation). Here in the
former case the recombined $s\bar s$ in the final state can be
projected on the $\eta$ meson according to the following relations
(see,e.g.\cite{chao})
 \bqa
 s\bar {s}&=& 1/\sqrt{3}(cos\theta-\sqrt{2}sin\theta)\eta'
 -1/\sqrt{3}(\sqrt{2}cos\theta+sin\theta)\eta \nonumber \\
 &=& 0.72(0.82)\eta'-0.69(0.57)\eta,  \\
 1/\sqrt{2}(u\bar u+d\bar d)&=&1/\sqrt{3}(cos\theta-\sqrt{2}sin\theta)\eta
 +1/\sqrt{3}(\sqrt{2}cos\theta+sin\theta)\eta' \nonumber \\
 &=& 0.72(0.82)\eta+0.69(0.57)\eta',
\eqa
 where $\theta$ is the $\eta-\eta'$ mixing angle, and the
numerical values of the projection coefficients are obtained for
$\theta=-11^o(-20^o)$. From eq.(1) and the observed ratio
$R=\Gamma(D^0K^+)/\Gamma(D_{s}^{+} \eta)=0.16\pm 0.06$ we see that
the required suppression factor for the $D^0K^+$ decay mode would
be in fact larger than a factor of 13, indicating the demand that
the momentum in the $D^0K^+$ mode is very close to the zero of the
decay amplitude. As already noted that in general the zeros in the
decay amplitude are very sensitive to the wave functions and
interquark potentials in use \cite{chaoding}. So, obviously a
quantitative understanding for the $D_{SJ}^+(2632)$ (as the $2S$
radial excitation) decay widths of various channels would need an
elaborate model to perform the calculations and to see whether the
required suppression can be realized. This will be left for the
future consideration. (Note that the decay to the P-wave charmed
meson e.g. $D_1(2420) (J^P=1^+)$ associated with the kaon would be
favored for the $D_{SJ}^+(2632)$ decay if it is the $J^P=1^-$
radial excitation, but it is kinematically forbidden.)

If the $D_{SJ}^+(2632)$ is the first radial excitation of the
$1^-$ ground state $\rm{D_s}^*$(2112), it would also decay to
$\rm{D_s}^*(2112)\pi\pi$ but only with a small branching ratio.
The radiative decay to $\rm{D_s}(1969)\gamma$ should be totally
negligible. The quantum number $J^P=1^-$ could be examined from
the angular distributions of the decay products. Moreover, we
should also see the first radial excitation of the $0^-$ ground
state $\rm{D_s}$(1969) at about 2490~MeV, of which all the OZI
allowed decays e.g. $D^*K,DK^*,D^*K^*,...$ are kinematically
forbidden, and it can only decay to $D_s\pi\pi$ via soft gluon
emition and hadronization to $\pi\pi$. Of course, if its mass is
above the $D^*K$ threshold it would have the OZI allowed decays.
The first radial excitation of the $\rm{D_s}^*$(2112) should be
observable in the $B$ meson decays at Belle and BaBar, and in the
$e^+e^-$ annihilation at BES and CLEO-c, and may also be seen from
the $e^+e^-$ continuum at $\sqrt{s}=10.6~GeV$ at Belle and BaBar,
where the charmed-nonstrange meson pair production processes
$e^+e^-\rightarrow D^{(*)}\bar D^{(*)}$ have already been observed
by Belle\cite{uglov}, and it should not be too difficult to detect
$D_s^{(*)}$ mesons and their radial excitations with higher
statistics in the near future.

We finally note that any orbitally excited states of the $c\bar s$
mesons (without nodes in their wave functions) such as the D-wave
$l=2, j=3/2,5/2$ light quark excitations with
$J^P=1^-,2^-,2^-,3^-$ for the $c\bar s$ systems could hardly
explain the unusual decay patten that
$R=\Gamma(D^0K^+)/\Gamma(D_{s}^{+} \eta)=0.16\pm 0.06$.

\vspace{0.5cm} (2). The $D_{SJ}^+(2632)$ could be a $c\bar sg$
hybrid state. \vspace{0.5cm}

In some constituent gluon model for the hybrid states (see
e.g.\cite{m,chao}) the decay may proceed via the gluon conversion
into the color-octet quark pairs (assuming SU(3) symmetry) \be
g\rightarrow 1/\sqrt{3}(u\bar u+d\bar d+s\bar s)_8, \ee and then
the color-octet quark pairs become the neutralized color-singlet
ones by the gluon exchange with the color-octet $(c\bar s)_8$
component. If the gluon exchange does not flip the quark spin
(e.g. via the long-ranged color-electric force), the final state
from the gluon conversion in the hybrid would become the $\omega$
or $\phi$ mesons. If the short-ranged magnetic color-spin force
induced by one gluon exchange flips the quark spin, the final
state from the gluon in the hybrid would become the $\eta$ or
$\eta'$ mesons. The short-ranged magnetic color-spin force could
make the hybrid narrow. Hence a $c\bar sg$ hybrid could mainly
decay to a $D_s$ meson plus a $\eta$ meson if its mass does not
allow it decaying into $\omega$ or $\phi$ mesons in the final
state. The $c\bar sg$ hybrid state could also decay into $DK$
mesons via the quark rearrangement from the $(c\bar s)_8(u\bar
u+d\bar d+s\bar s)_8$ configuration. It is not clear dynamically
whether the quark rearrangement can compete with the gluon
exchange. If not, then the ${D_{s}\eta}$ decay mode would
dominate. However, this is still an open question in the
constituent gluon model for the hybrid states because of the
complication of the nonperturbative dynamics.

A study for the heavy-light hybrid $Q\bar qg$ in the QCD sum rule
approach may further shed light on the masses and decay widths of
the $Q\bar qg$ states\cite{zhu}. It shows that in the heavy quark
limit the lowest lying $Q\bar sg$ state is the $J^P=1^-$ state
with the strange quark and gluon excitation energy of about 1.7
GeV, and the $J^P=1^+,0^+,0^-$ hybrids will have higher masses.
The total decay width of the $1^-$~$Q\bar qg$ state would be about
300~MeV, but more than $80\%$ are due to decays into the P-wave
charmed meson associated with a light pseudoscalar meson (i.e. the
partial decay width to the ground state charmed mesons is only
about 10 ~MeV). If the observed $D_{SJ}^+(2632)$ is the $1^-$~
$c\bar sg$ state (the mass estimate in the heavy quark limit for
the charmed hybrid might suffer from large $1/m_c$ corrections),
it would then have a narrow width of the order 10~MeV, since
decays involving the P-wave charmed mesons would not be
kinematically allowed. This would seem to be encouraging. But the
mass of this $c\bar sg$ hybrid state is estimated to be higher
than 3.0 GeV, and it is hard to lower its mass down to 2.63 GeV by
$1/m_c$ corrections. Therefore it is not very likely to have this
hybrid state as the candidate of $D_{SJ}^+(2632)$.

\vspace{0.5cm} (3). The $D_{SJ}^+(2632)$ could be a charmed
baryonium (diquark-antidiquark) $(cs)_{3^*}-(\bar s\bar s)_{3}$
state. \vspace{0.5cm}

In the recent studies for the pentaquarks, the diquark correlation
has been emphasized \cite{jw}. If the diquark correlation does
exist even in the S-wave hadrons, the diquark-antidiquark bound
state or resonance (it was sometimes called the baryonium since it
would easily decay to a baryon and antibaryon pair via a light
quark pair creation if it has enough phase space) should also
exist. As examples the $(cq)_{3^*}-(\bar c\bar q)_{3}$ (see,
e.g.\cite{chao1}) and $(sq)_{3^*}-(\bar s\bar q)_{3}$ (see, e.g.
\cite{chao}) states have been discussed along with lots of studies
by other authors. Here we would like to suggest the
$(cs)_{3^*}-(\bar s\bar s)_{3}$ state be a possible candidate for
the observed $D_{SJ}^+(2632)$ state (see also \cite{lm,cl} for
discussions on the four quark states).  As in \cite{jw}, assuming
the diquark correlation exists even in the S-wave hadrons (this is
different from the scenario of diquark clusters which are
separated by large angular momentum barriers), then the
$(cs)_{3^*}-(\bar s\bar s)_{3}$ state can only decay via the quark
rearrangement into the $c\bar s$ and $s\bar s$ ($D_s\eta$) mesons
and then may have a narrow width. The $s\bar s$ can further mix
with $u\bar u+d\bar d$ and this small mixing (not far from the
ideal mixing) would lead to a small decay branching ratio to the
$DK$ mesons. Here the key assumption is that the $(cs)_{3^*}-(\bar
s\bar s)_{3}$ has very small overlap with the $(c\bar s)_1-(s\bar
s)_1$ color-configuration due to the diquark correlation, so that
the $(cs)_{3^*}-(\bar s\bar s)_{3}$ can not simply fall apart into
the $D_s\eta$ mesons with a broad width (even for the S-wave
state). If this is a right picture, we would have a nice
interpretation for the $D_{SJ}^+(2632)$. But we have to know the
dynamics for the diquark correlation. This is certainly a very
interesting subject in low energy QCD. Here the $\bar s\bar s$
diquark must have spin one since both flavor and spin are
symmetric under the exchange of two quarks. Therefore the lowest
$(cs)_{3^*}-(\bar s\bar s)_{3}$ state could have $J^P=1^+$.

In summary, we have suggested some possible interpretations for
the $D_{SJ}^+(2632)$ observed by SELEX. The $D_{SJ}^+(2632)$ could
be the first radial excitation of the $1^-$ ground state
$\rm{D_s}^*$(2112), and its unusual decay patten might be
hopefully explained by the node structure of the wave functions.
In addition, the $D_{SJ}^+(2632)$ could also be a $c\bar sg$
hybrid state or a $(cs)_{3^*}-(\bar s\bar s)_{3}$
(diquark-antidiquark) state, but these two assignments are less
likely than the first one.

This work was supported in part by the National Natural Science
Foundation of China, the Education Ministry of China, and the
Beijing Electron Positron Collider National Lab.

$Note~~ added.$ After this work appeared in arXiv:hep-ph/0407091,
Barnes et al.\cite{barnes} and van Beveren et al.\cite{beveren}
independently reached the same conclusion that the
$D_{SJ}^+(2632)$ could be the first radial excitation of the $1^-$
ground state $\rm{D_s}^*$(2112).

\end{document}